\documentclass[aps,pre,preprint,superscriptaddress,amsmath,amssymb]{revtex4-2}
\usepackage{graphicx}
\usepackage{dcolumn}
\usepackage{bm}
\usepackage[mathlines]{lineno}
\usepackage{epsfig}
\usepackage{epstopdf}
\usepackage{amsmath,amssymb,amsthm}
\usepackage{bm}
\usepackage{float}
\usepackage{xcolor}
\usepackage{mathtools}
\usepackage{subfig}
\usepackage{algcompatible}
\usepackage{newfloat}
\DeclareFloatingEnvironment[
fileext=loa,
listname=List of Algorithms,
name=ALGORITHM,
placement=tbhp,
]{algorithm}

\begin{document}
\title{Approximations of the modified Bessel functions of the second kind $K_\nu$. Applications in random field generation.}
	\author{D. I. Palade}
	\email{dragos.palade@inflpr.ro}
	\author{L. Pom\^{a}rjanschi}
\email{ligia.pomarjanschi@inflpr.ro}
\affiliation{National Institute of Laser, Plasma and Radiation Physics,	M\u{a}gurele, Romania }
\affiliation{Faculty of Physics, University of Bucharest, M\u{a}gurele, Romania
}

\date{\today}

\begin{abstract}
We propose an analytical approximation for the modified Bessel function of the second kind $K_\nu$. The approximation is derived from an exponential ansatz imposing global constrains. It yields local and global errors of less than one percent and a speed-up in the computing time of $3$ orders in magnitude in comparison with traditional approaches. We demonstrate the validity of our approximation for the task of generating long-range correlated random fields.
\end{abstract}

	\keywords{modified Bessel, approximation, random field}

\maketitle
\section{Introduction}
\label{intro}
Bessel functions \cite{watson1995treatise,bowman2012introduction} represent a special class of mathematical functions in the theory of differential equations, analytic numbers and integral transforms with applications in many branches of natural sciences \cite{korenev2002bessel}. A variety of accurate approximations have been developed for special subclasses, such as $J_\nu(x)$ \cite{blachman1986trigonometric,harrison2009fast,krasikov2014approximations,li2006new,millane2003polynomial}. For the modified Bessel functions of the second kind, $I_\nu(z), K_\nu(z)$, only particular approximations are available: $I_0(x)$ \cite{olivares2018simple},  $I_1(x)$ \cite{martin2021quasi,martin2017precise}, $K_0(x)$ \cite{caruso2021new}, etc. 

$K_\nu(x)$ has many applications in the physics of relativistic ideal gases \cite{louis2011classical}, vibrating membranes, waveguides \cite{lee2007wave}, etc. $K_\nu(x)$ is also involved in the spectrum of random fields with long-range correlation functions. The latter arise naturally in the study of turbulence \cite{PhysRevLett.80.4438}, biology\cite{rangarajan2003processes}, finance \cite{COSTA2003231} , etc. and are an indicator of highly coherent, collective phenomena, which might exhibit self-organization or self-similarity. 

We propose, as the central result of this work, a simple approximation $K_\nu^{app}(x)$ for the modified Bessel function of the second kind $K_\nu(x), \forall \nu > 0, x > 0$: 

\begin{align}
	\label{eq_0}
	\begin{cases}
		K^{app}_\nu(x) = e^{-\left(\frac{x}{\lambda(\nu)}\right)^{\gamma(\nu)}}x^{-\nu}\Gamma(\nu)2^{\nu-1}\\
		\lambda(\nu) = \gamma(\nu)\sqrt{\pi}\frac{\Gamma(\nu+1/2)}{\Gamma(\nu)\Gamma(1/\gamma(\nu))}\\
		\gamma(\nu) = 2\frac{(2\nu)^{c(\nu)}}{1+(2\nu)^{c(\nu)}}\\
		c(\nu) = \frac{0.2168 + 0.932 \nu}{0.392 + \nu}
	\end{cases}
\end{align}
where $\Gamma(x)$ is the standard Gamma function.

The paper is structured as it follows. In the Derivation section (\ref{section_1}) we show how eqns. \ref{eq_0} were obtained and analyse their strength and limitations. In the Application section (\ref{section_2}) the approximation \ref{eq_0} is used to generate long-range correlated random fields. This application was chosen in order to asses how the errors induced through the approximation propagate to complex tasks such as the construction of numerical stochastic objects, i.e. random fields.

\section{Derivation}
\label{sec:1}
In order to derive eq. \ref{eq_0}, we shall work further with a \emph{normalized} version of the modified Bessel function of the second kind $K_\nu(x)$ denoted $\tilde{K}_\nu(x)$ and defined as:

\begin{align}
	\label{eq_1}
	\tilde{K}_\nu(x) = \frac{K_\nu(x)x^\nu}{2^{\nu-1}\Gamma(\nu)}
\end{align}
The reason for this choice is two-fold. First, $\tilde{K}_\nu(x)$ is finite near the origin $\tilde{K}_\nu(0)=1$ (in contrast with  $K_\nu$ which is divergent) thus, easier to approximate. Secondly, as we shall see in section \ref{sec:2}, this form is intimately related to long-range correlation functions. 
$\tilde{K}_\nu(x)$ is a monotonically decreasing function which decays asymptotically as $\tilde{K}_\nu(x) \sim e^{-x}x^{\nu-1/2}$. It has analytical expressions $\forall x>0$ when $2\nu-1\in \mathbb{Z}$. In particular, the relation $\tilde{K}_{1/2}(x) = e^{-x}$ holds true. We propose as ansatz that $\tilde{K}_\nu(x)$ can be approximated with an exponentially decaying function $\tilde{K}^{app}_\nu(x) = \exp\{-(x/\lambda(\nu))^{\gamma(\nu)}\}$.

Further, we have to estimate the functions $\lambda(\nu),\gamma(\nu)$ that make $\tilde{K}^{app}_\nu(x)$ a good fit for the real $\tilde{K}_\nu(x)$. This is achieved by imposing two global constrains on the ansatz: the first two moments of $\tilde{K}_\nu(x)$ are well reproduced by $\tilde{K}_\nu^{app}(x)$, i.e. $m_\alpha=m_\alpha^{app}, \hspace*{0.1cm} \forall \alpha\in\{0,1\}$, where the moments are defined as:

\begin{eqnarray}
	\label{eq_2}
	m_\alpha^{(app)} =  \int_0^\infty\tilde{K}_\nu^{(app)}(x)x^\alpha ~dx 
\end{eqnarray}   

These integrals can be computed analytically: 

\begin{align}
	\label{eq_3}
	m_\alpha &= \frac{2^\alpha \Gamma\left(\frac{\alpha+1}{2}\right)\Gamma\left(\nu+\frac{\alpha+1}{2}\right)}{\Gamma\left(\nu\right)}\\
	m_\alpha^{app} &= \Gamma\left(\frac{1+\alpha}{\gamma(\nu)}\right)\frac{\lambda(\nu)^{1+\alpha}}{\gamma(\nu)}
\end{align}

From the condition $m_0^{app} = m_0$ we recover the expression for $\lambda(\nu)$ in terms of $\gamma(\nu)$:

\begin{align}
	\label{eq_4}
	\lambda(\nu) = \frac{\sqrt{\pi } \gamma(\nu)  \Gamma \left(\nu +\frac{1}{2}\right)}{\Gamma\left(1/\gamma(\nu)\right) \Gamma (\nu )}
\end{align}

The second condition $m_1^{app} = m_1$ can be solved only numerically to obtain the $\gamma(\nu)$ function. Doing that, we have found that $\gamma(\nu)$ is a monotonically increasing function that starts as $\gamma(\nu\to 0 ) = 0$, grows and saturates at $\gamma(\nu\to\infty) = 2$. An imperative condition considered was that $\gamma(1/2) = 1$, in order to recover the exact case $\tilde{K}_{1/2}^{app}(x) = e^{-x} = \tilde{K}_{1/2}(x) $. For these reasons, we searched for a quasi-rational estimation of the $\gamma$ exponent that obeys the last requirements:

\begin{align}
	\label{eq_5}
	\gamma(\nu) \approx 2\frac{\left(2\nu\right)^c}{1+\left(2\nu\right)^c}
\end{align}

It turns out that this shape is a good fit for the numerical values of $\gamma(\nu)$ only if we allow for some dependencies $c(\nu)$. Using a numerical fitting procedure, we find the last result of eqns. \ref{eq_0}:

\begin{align}
	\label{eq_6}
	c(\nu) \approx \frac{0.2168 + 0.932 \nu}{0.392 + \nu}
\end{align}

Now that all elements of the main result of this work \ref{eq_0} have been derived, it is important to asses its validity in a quantitative manner. First, the approximation is exact, by construction, for $\nu = 1/2$ at all $x\in \mathbb{R}$. For $\nu\neq 1/2$ the approximation is exact in the limit of $x\to 0$ and fails at $x\to\infty$ where the asymptotic correct behaviour $\sim x^{-1/2}e^{-x}$ is approximated by $\sim x^{-\nu}e^{-(x/\lambda)^{\gamma(\nu)}}$. 

\begin{figure}
	\subfloat[\label{fig_1a}]{
		\includegraphics[width=.50\linewidth]{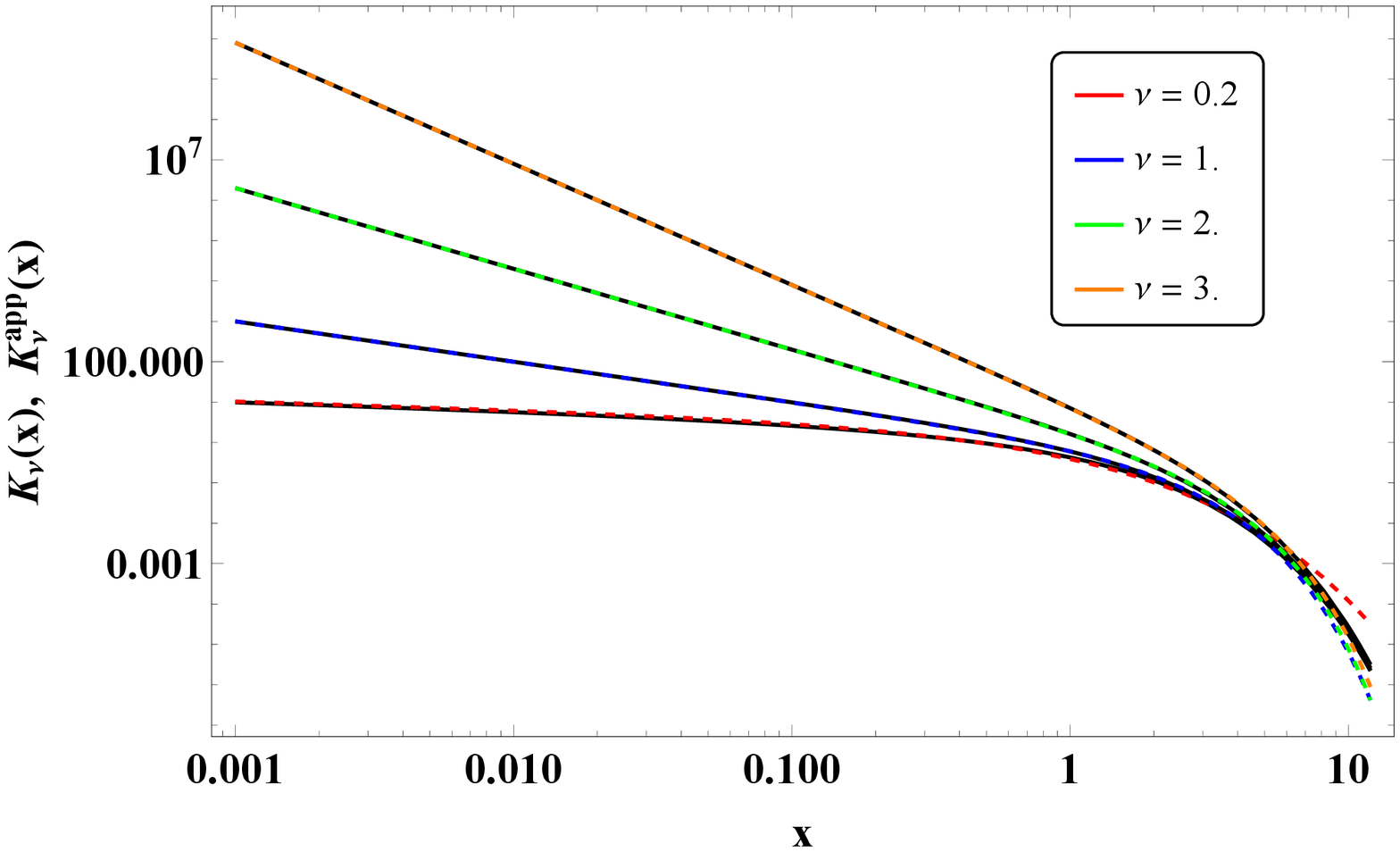}%
	}	\hspace*{0.5cm}
	\subfloat[\label{fig_1b}]{%
		\includegraphics[width=.48\linewidth]{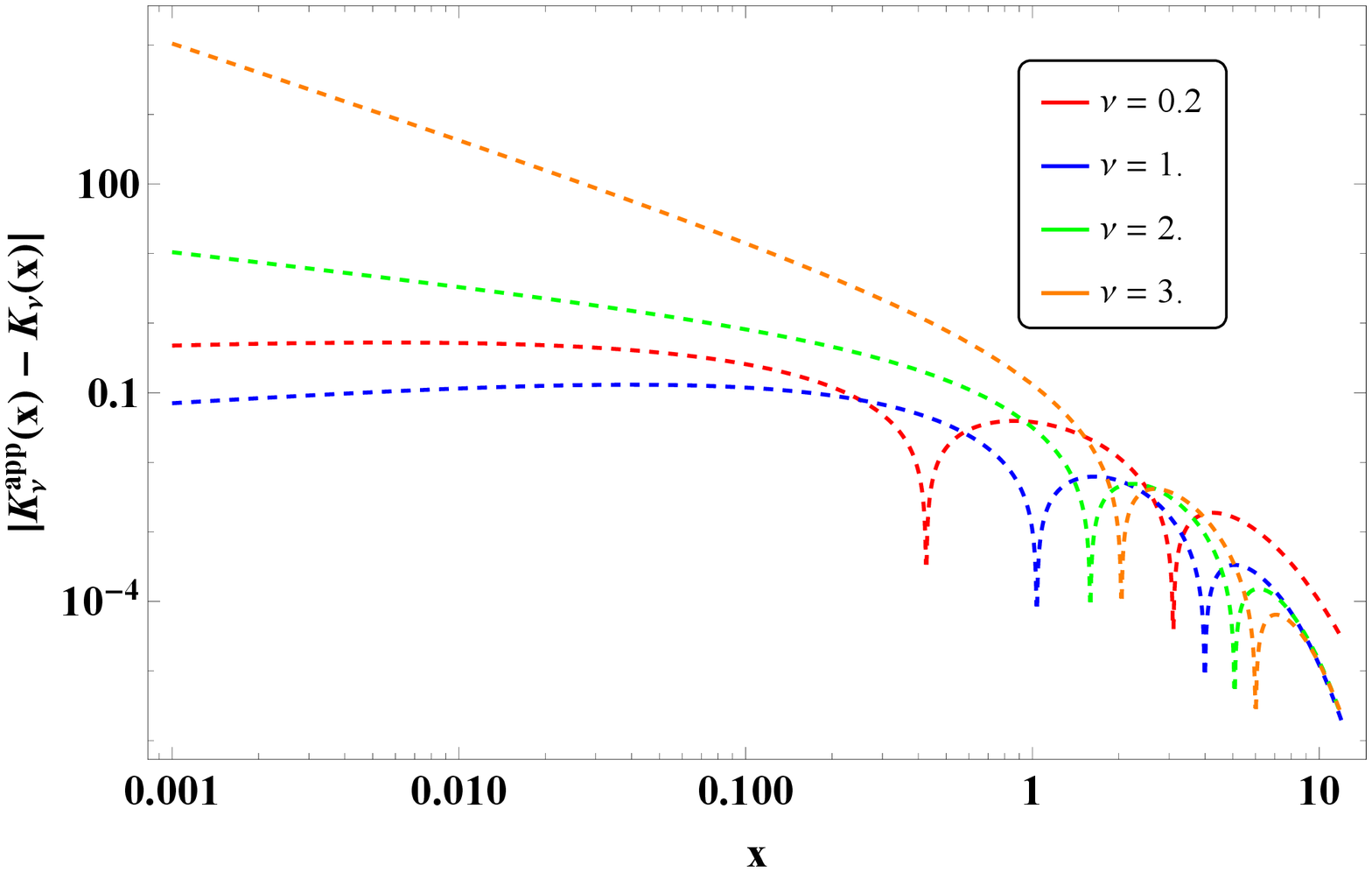}%
	}	\\
	\subfloat[\label{fig_1c}]{%
		\includegraphics[width=.49\linewidth]{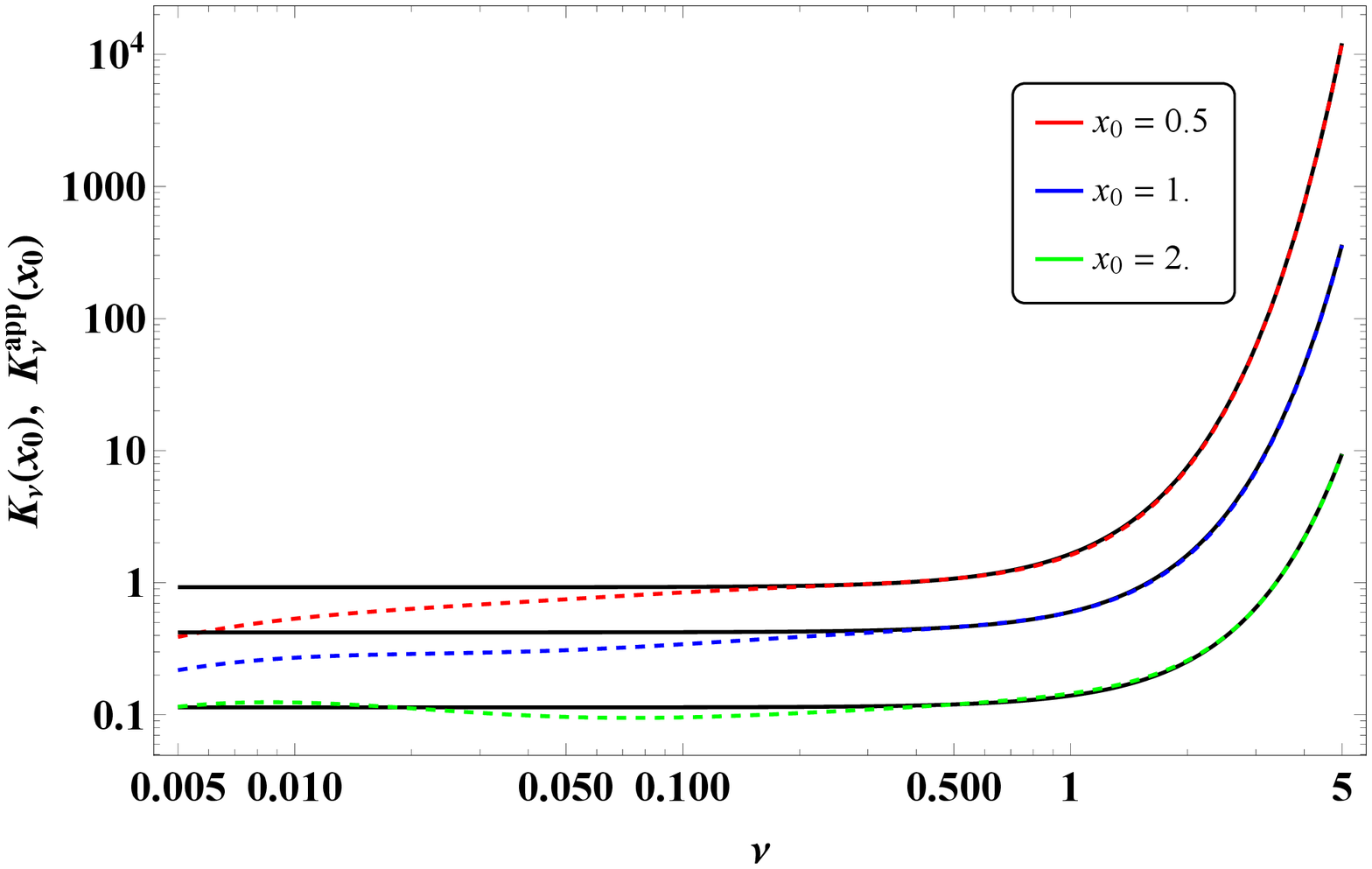}%
	}	
	\caption{(a) The exact $K_\nu(x)$ (black lines) vs $K_\nu^{app}(x)$ (dashed, coloured, lines). (b) the local absolute error $\mid K_\nu^{app}(x)-K_\nu(x)\mid$ of the approximation is shown. c)  $K_\nu(x)$ (black lines) vs $K_\nu^{app}(x)$ as functions of $\nu$ at different fixed $x_0$. Each value of the parameter $\nu\in\{0.2,1.0,2.0,3.0\}$ or $x_0\in\{0.5,1.0,2.0\}$ is colour coded.}
\end{figure}

In Fig. \ref{fig_1a} we show in log-log scale the exact $K_\nu(x)$ function and its approximate value $K^{app}_\nu(x)$. As expected, the dashed lines (approximation) breaks down at large $x$ (departs from the black lines). 
In Fig. \ref{fig_1b} we show the local errors $\mid K_\nu(x)-K^{app}_\nu(x)\mid$ for several $\nu$ values. While they seem quite large at $x\sim 10^{-2}$ and lower, they are just $\sim 1\%$ of the $K_\nu(x)$. Fig. \ref{fig_1c} indicates that $K^{app}_\nu(x)$ breaks down at small $\nu$, especially $\nu<1/2$. 

\begin{figure*}
	\subfloat[\label{fig_2a}]{
		\includegraphics[width=.49\linewidth]{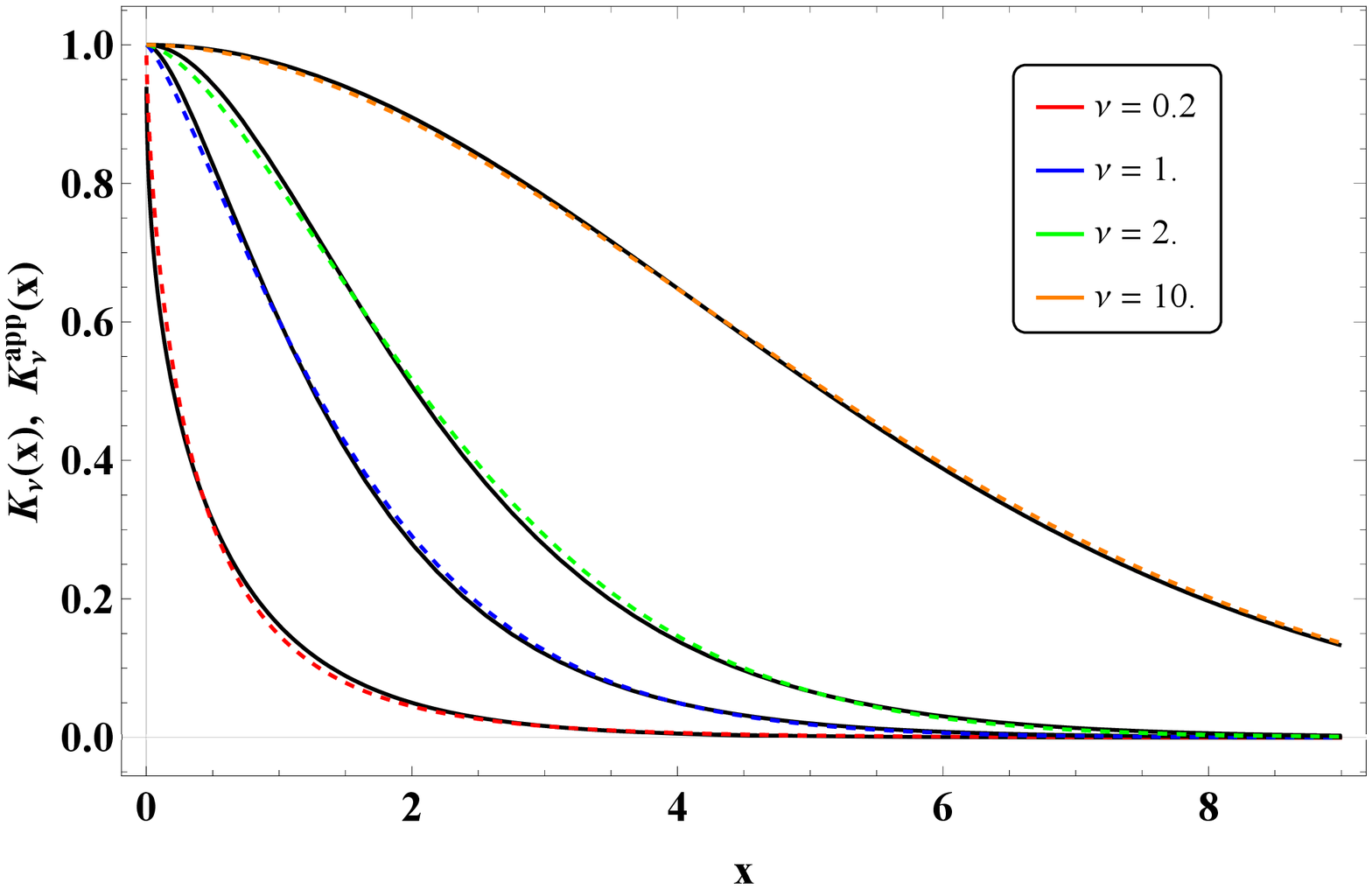}%
	}	\hspace*{0.5cm}
	\subfloat[\label{fig_2b}]{%
		\includegraphics[width=.49\linewidth]{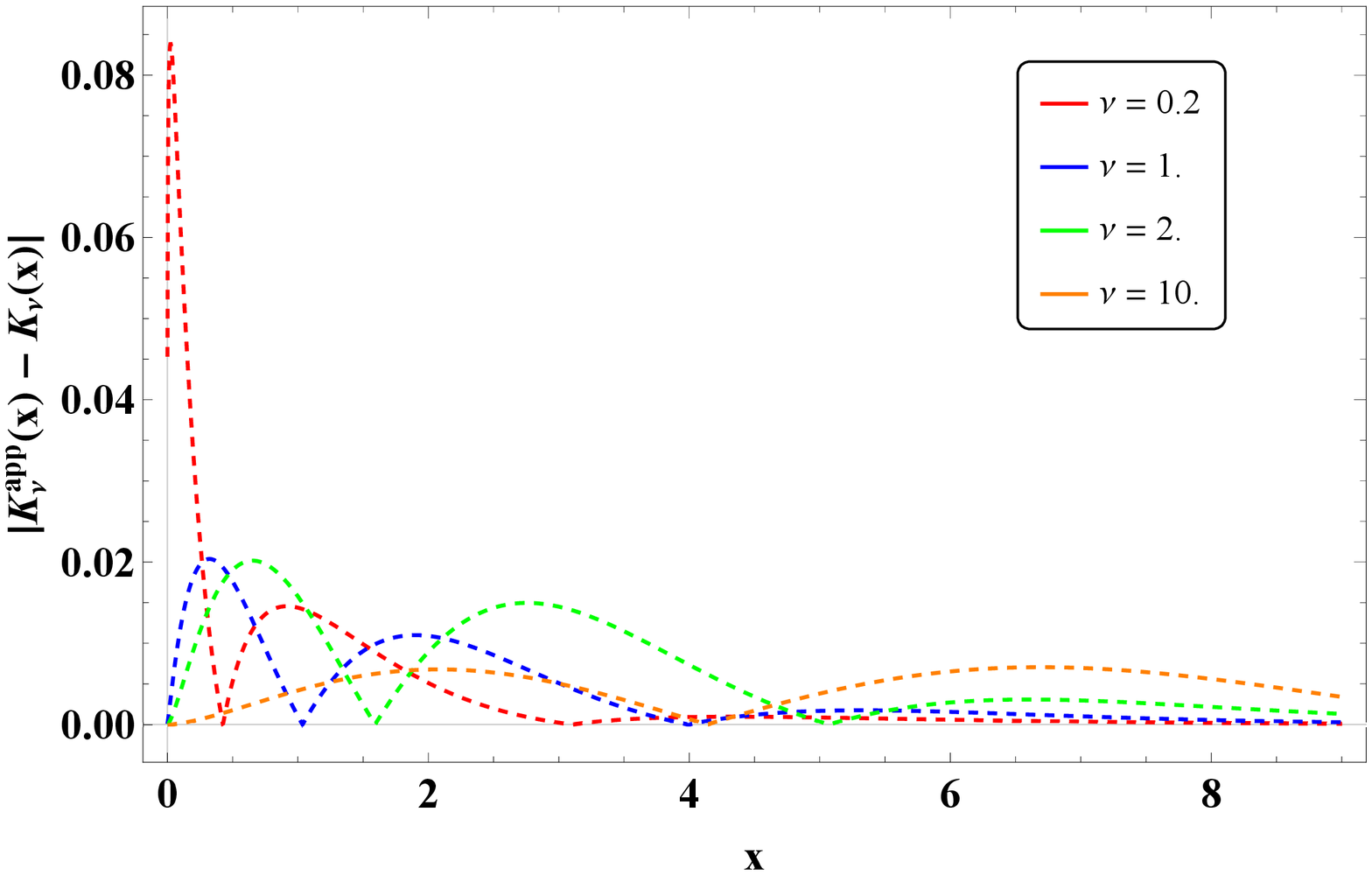}%
	}	
	\caption{The exact $\tilde{K}_\nu(x)$ (full line) and the approximate $f_\nu(x)$ (dashed line) profiles plotted at different $\nu$ values (colour encoded). In Fig. b), the local errors $\mid\tilde{K}_\nu^{app}(x)-\tilde{K}_\nu(x)\mid$ are shown.}
\end{figure*}

Another way of looking at the validity of our approximation is to compare the exact $\tilde{K}_\nu(x)$ vs. $\tilde{K}^{app}_\nu(x)$. We do that in Figs. \ref{fig_2a}, \ref{fig_2b} where the functions are shown against each other and their local error. Again, our expectation is met with larger errors for low $\nu$ and close to $x\to 0$. 

\begin{figure}
	\centering
	\includegraphics[width=.89\linewidth]{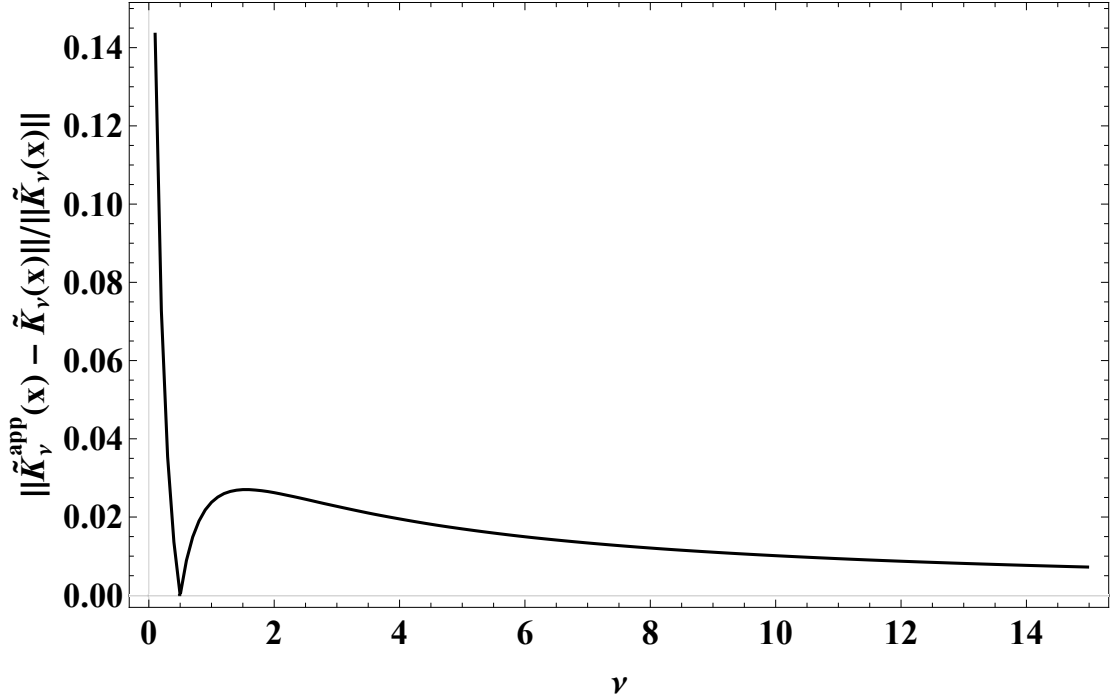}%
	\caption{Local errors of the approximation $f_\nu(x)$ at different $\nu$ values.}
	\label{fig_3a}	
\end{figure}

Finally, we evaluate the global error which we define it as $\|\tilde{K}_\nu^{app}(x)-\tilde{K}_\nu(x)\|/\|\tilde{K}_\nu(x)\|$, where $\|f(x)\| = \int_0^\infty |f(x)| ~dx$. The result is shown in Fig. \ref{fig_3a} where one can see that for $\nu=1/2$ the global error is $0$ while for $\nu>1/2$ the error stays $\sim 1-2\%$. The approximation breaks down fast for $\nu<1/2$ where we find errors as large as $\sim 10\%$.  

We have compared the CPU computing time for our approximation vs standard implemented function in Mathematica Wolfram \cite{wolfram1999mathematica}. For large chunks of data, the use of formula \ref{eq_0} improves the overall computing time with $3-4$ orders of magnitude. Obviously, for half-integer $\nu$, the Bessel function takes simple, analytical, forms which are easily computed by any compiler. We conclude at this end that, when involving large sets of data, the gain in CPU time can justify the use of the approximation \ref{eq_0} despite the relatively large errors $\sim 1\%$.

\section{Applications in random field generation}
\label{sec:2}

The modified Bessel functions of the second kind have many applications in the field of natural sciences. In particular, in statistics, various random variables of interest have distributions related to $K_\nu(x)$ or $I_\nu(x)$. We consider in this section the problem of generating Gaussian random fields with long-range correlation, which, as we shall see, are intimately related to $K_\nu(x)$. The task will be tackled using \ref{eq_0} and the validity of the results will be analysed.

The reason behind choosing such a particular application is the following: while the gross errors induced by our approximation have been evaluated in the previous section it is important to see how this errors tend to propagate in more complex tasks. The generation of long-range-correlated Gaussian random fields is a good example of such a test.

Let there be a real-valued Gaussian random field $\phi(\mathbf{x}), \mathbf{x}\in\mathbb{R}^n$. We consider that this field is characterized by a zero statistical average $\langle\phi(\mathbf{x})\rangle = 0, \forall\mathbf{x}$ and the two-point correlation function $\langle \phi(\mathbf{x})\phi(\mathbf{y})\rangle = \mathcal{E}(\mathbf{x};\mathbf{y})$. For simplicity, we ask that the field is homogeneous, i.e. $\mathcal{E}(\mathbf{x};\mathbf{y}) = \mathcal{E}(\mid\mathbf{x}-\mathbf{y}\mid)$. While there are many ways to construct numerically such objects \cite{Liu2019,Palade2021} we choose here to use a Fourier-series-like decomposition:
\begin{align}\label{eq_12}
	\phi(\mathbf{x}) = \sqrt{\frac{2}{N_c}}\sum_{i=1}^{N_c} \sin\left(\mathbf{k}_i\mathbf{x}+\alpha_i\right)
\end{align}
where $\alpha_i$ are independent and identically distributed random variables (iidrv), uniform in $(0,2\pi)$. The vectors $\mathbf{k}_i$ are also iidrv with their PDF described by the function $S(\mathbf{k}) = \int \mathcal{E}(\mathbf{x})e^{i\mathbf{k}\mathbf{x}}~d\mathbf{x}$. If these conditions are met, one can show that an ensemble of fields constructed via eq. \ref{eq_12} is indeed zero-average and has the correct correlation function $\mathcal{E}(\mid\mathbf{x}-\mathbf{y}\mid)$. If the limit $N_c\to\infty$ is met, the field is also Gaussian (implied by the central limit theorem). Thus, the main task in GRF generation via eq. \ref{eq_12}, is to generate random variables $\mathbf{k}_i$ with the appropriate PDF $S(\mathbf{k})$. 

We choose here two classes of long-range correlation functions. The first is denoted $\mathcal{E}_1(t)$ and it is designed for time dependent fields $\phi(t)$ displaying a behaviour similar to the pdf of the Student's T distribution. The second is denoted as $\mathcal{E}_2(x,y)\equiv \mathcal{E}_2(r), r=\sqrt{x^2+y^2}$ for a two-dimensional fields which displays both long-range and oscillating character. We provide  their explicit expressions and the associated spectra $S$ (the Fourier transform of the correlation): 

\begin{align}
	\label{eq_11}
	\mathcal{E}_1(t) = \left(1+\frac{t^2}{\nu}\right)^{-\frac{\nu+1}{2}} &\implies S_1(\omega) \propto \mid\omega\mid^{\nu/2}K_{\nu/2}(\mid\omega\mid\sqrt{\nu}) \\
	\mathcal{E}_2(r) = {}_2 F_1\left(\frac{3}{2},\frac{1}{2}+\nu,1,-\frac{r^2}{2\nu}\right) &\implies S_2(k) \propto k^{\nu+1} K_{\nu-1}(k\sqrt{2\nu})
\end{align}
where ${}_2 F_1$ is the ordinary hypergeometric function while $k=\sqrt{k_x^2+k_y^2}$. 

We proceed further to generate random numbers accordingly with the spectra $S_1(\omega)$, $S_2(k)$. In order to do that we employ the acceptance-rejection method \cite{doi:10.1137/1032082}. The Bessel functions are evaluated with the approximation \ref{eq_0}. A number of $N_c\times N_p$ random variables $\omega_i,\mathbf{k}_i$ have been generated, where $N_c = 100$ and  $N_p = 10^6$ the dimension of the statistical ensemble. Consequently, a number of $N_p$ fields $\phi(t),\phi(x,y)$ have been obtained.

\begin{figure*}
	\subfloat[\label{fig_4a}]{
		\includegraphics[width=.49\linewidth]{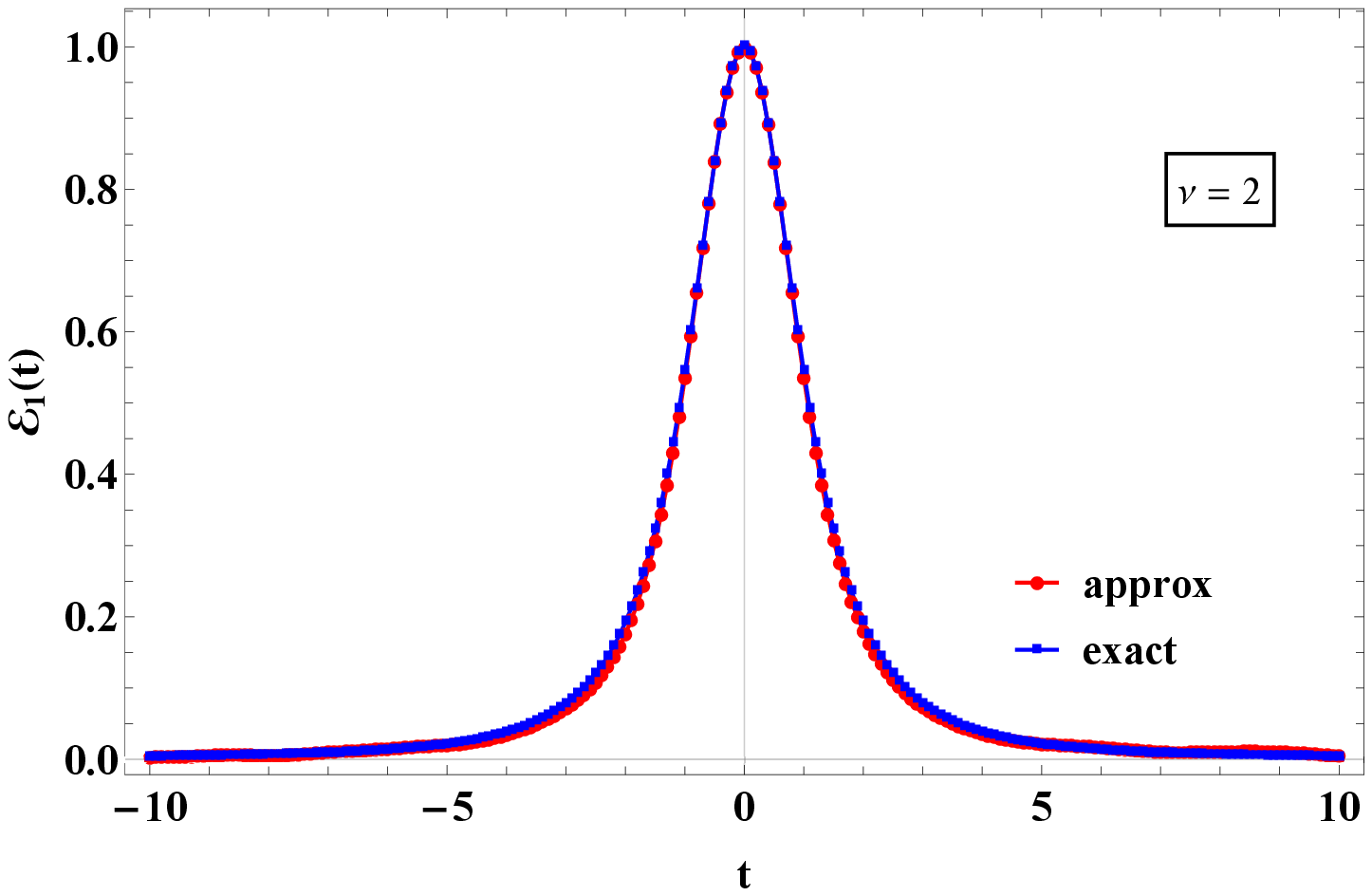}%
	}	
	\subfloat[\label{fig_4b}]{%
		\includegraphics[width=.49\linewidth]{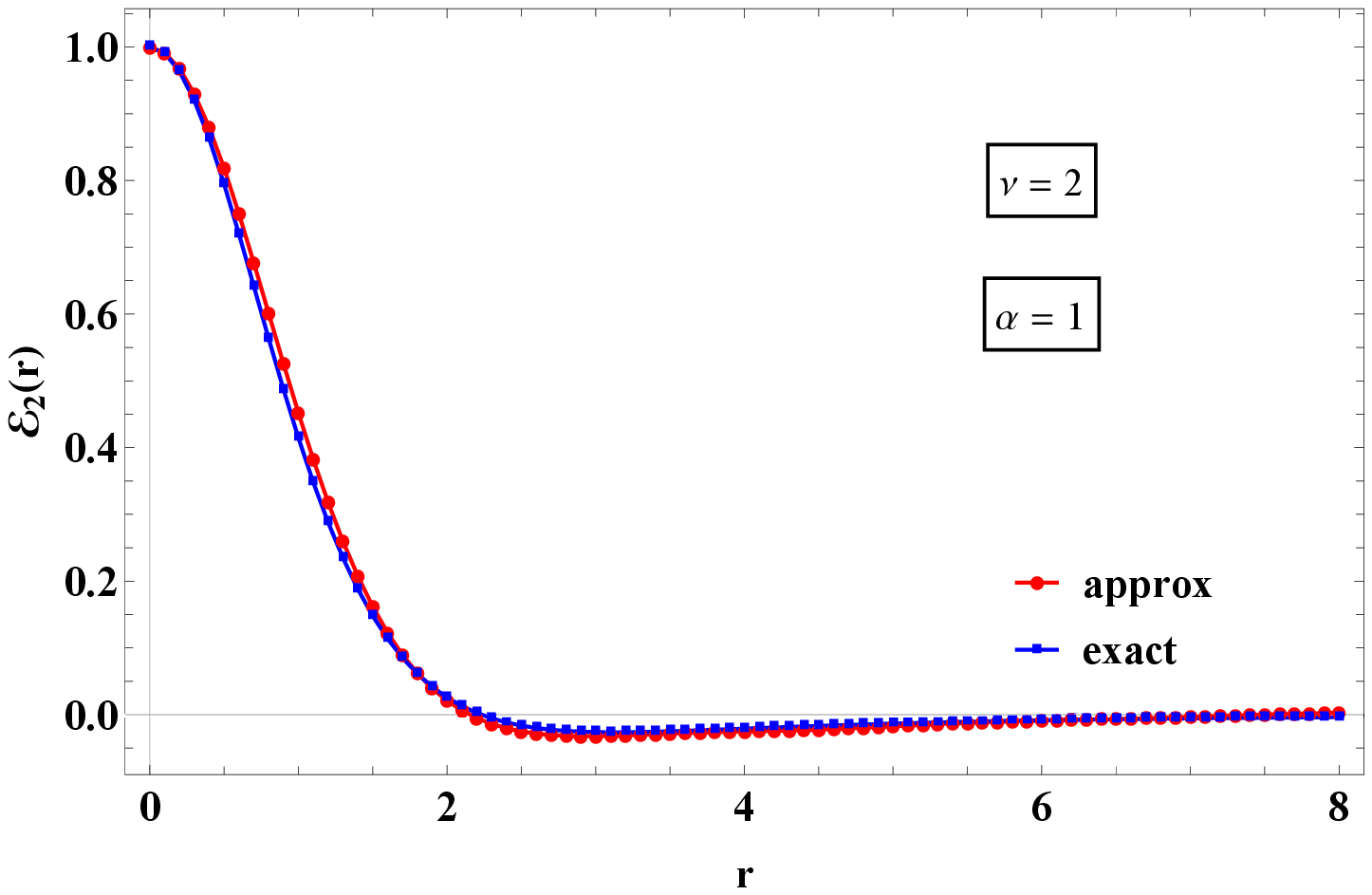}%
	}	
	\caption{Exact (blue, squared markers) correlations $\mathcal{E}_1$ (left, a)) and $\mathcal{E}_2$ (right, b)) vs. the numerical results (red, round markers) obtained from the random fields generated with the aid of approximation \ref{eq_0}.}
\end{figure*}

In Fig. \ref{fig_4a} we show the exact correlation function $\mathcal{E}_1(t)$  (blue, squared markers) vs. the numerical correlation function obtained through the averaging of $\langle\phi(0)\phi(t)\rangle$ (red, round markers). The value $\nu = 2$ was employed. A similar comparison is done in Fig. \ref{fig_4b} where the exact correlation with radial symmetry $\mathcal{E}_2(r)\equiv \mathcal{E}_2(x,y)$ is compared with the numerical correlation resulting from $\langle\phi(0,0)\phi(0,\sqrt{x^2+y^2}=r)\rangle$. The value of $\alpha = 1$ was used. As one can see, the statistical features of the generated random fields, i.e. their correlation, fits quite well the exact, analytical values. This is an indicator that our approximation is relatively valid even for more complex tasks involving the evaluation of modified Bessel functions of the second kind $K_\nu(x)$.

\section{Conclusions}

In this work we have presented an analytical approximation for the modified Bessel function of the second kind $K_\nu(x)$ which is valid at $x\ge 0, \nu>0$. The approximation is exact for $\nu=1/2$ and works well for $\nu>1/2$ and relatively small $x < 3\nu^{1/2}$ with local and global errors in the range $\sim 1\%$. It breaks down for small $\nu\ll 1/2$ where the errors grow up to $\sim 10\%$. 

The validity of the present approximation is proven in the task of generating homogeneous Gaussian random fields with long-range correlation. It is shown that the propagation of errors within such a complex task remains in the same relative range of errors, namely $\sim 1\%$. 

While the errors may be considered large for various cases (small $\nu$), the present result \ref{eq_0} is a good alternative for the traditional numerical methods of computing modified Bessel functions due to the gain in CPU time which is roughly that of $3-4$ orders in magnitude.

\section*{Acknowledgement}
	This research was partially supported by Romanian Ministry of Research, Innovation and Digitalization under Romanian National Core Program LAPLAS VII – contract no. 30N/2023.

\bibliographystyle{unsrt}
\bibliography{biblio}

\end{document}